\documentclass[fleqn,usenatbib]{mnras}

\usepackage{newtxtext,newtxmath}

\usepackage[T1]{fontenc}
\usepackage{ae,aecompl}

\usepackage{graphicx}	
\usepackage{amsmath}	
\usepackage{amssymb}	

\usepackage{color}
\definecolor{mygreen}{RGB}{84,141,40}

\title[A model for AGN variability]{A model for AGN variability on multiple timescales}

\author[Sartori et al.]{Lia F. Sartori$^{1}$\thanks{E-mail: lia.sartori@phys.ethz.ch}, Kevin Schawinski$^{1}$, Benny Trakhtenbrot$^{2}$, Neven Caplar$^{3}$, 
\newauthor
Ezequiel Treister$^{4}$, Michael J. Koss$^{5}$, C. Megan Urry$^{6}$ and Ce Zhang$^{7}$
\\
\\
$^{1}$Institute for Particle Physics and Astrophysics, ETH Z\"{u}rich, Wolfgang-Pauli-Str. 27, CH-8093 Z\"{u}rich, Switzerland, \\
$^{2}$Department of Physics, ETH Z\"{u}rich, Wolfgang-Pauli-Str. 27, CH-8093 Z\"{u}rich, Switzerland \\
$^{3}$Department of Astrophysical Sciences, Princeton University, Princeton, NJ 08544, USA \\
$^{4}$Instituto de Astrof\'{i}sica, Facultad de F\'{i}sica, Pontificia Universidad Cat\'{o}lica de Chile, Casilla 306, Santiago 22, Chile \\
$^{5}$Eureka Scientific Inc., 2452 Delmer St. Suite 100, Oakland, CA 94602, USA\\
$^{6}$Department of Physics, Yale University, P.O. Box 201820, New Haven, CT 06520-8120, USA\\
$^{7}$Systems Group, ETH Zurich, Universit\"{a}tstrasse 6, CH-8006 Z\"{u}rich, Switzerland
\\
}

\date{Accepted 2018 February 15}

\pubyear{2017}

\begin{document}
\label{firstpage}
\pagerange{\pageref{firstpage}--\pageref{lastpage}}
\maketitle

\begin{abstract}

We present a framework to link and describe AGN variability on a wide range of timescales, from days to billions of years. In particular, we concentrate on the AGN variability features related to changes in black hole fuelling and accretion rate. In our framework, the variability features observed in different AGN at different timescales may be explained as realisations of the same underlying statistical properties. In this context, we propose a model to simulate the evolution of AGN light curves with time based on the probability density function (PDF) and power spectral density (PSD) of the Eddington ratio ($L/L_{\rm Edd}$) distribution. Motivated by general galaxy population properties, we propose that the PDF may be inspired by the $L/L_{\rm Edd}$ distribution function (ERDF), and that a single (or limited number of) ERDF+PSD set may explain all observed variability features.
After outlining the framework and the model, we compile a set of variability measurements in terms of structure function (SF) and magnitude difference. We then combine the variability measurements on a SF plot ranging from days to Gyr. The proposed framework enables constraints on the underlying PSD and the ability to link AGN variability on different timescales, therefore providing new insights into AGN variability and black hole growth phenomena.


\end{abstract}




\section{Introduction}\label{sec:intro}

Variability in active galactic nuclei (AGN) is observed or inferred at all timescales, from hours to ${>}10^4\,{\rm yr}$. It is likely to be driven by physical processes originating at different spatial scales, from the nuclear region to the size of the galaxy.

Variability on days to decades timescales likely arises from accretion disc instabilities, mostly observed in the optical and UV bands. The AGN behaviour on these timescales has been studied through ensemble analysis (e.g. \citealt{Sesar2006}; \citealt{MacLeod2010}; \citealt{Caplar2017}). Years to decades timescales are observed also for the so-called ``changing look AGN'' or ``changing look quasars''\footnote{We note that the terms ``changing look AGN'' and ``changing look quasars'' are used inconsistently throughout the literature, and can refer to changes both in emission line width or obscuration, as well as in luminosity.}, where a change in AGN type is often accompanied by a change in luminosity (CLQSO, e.g. \citealt{LaMassa2015}; \cite{Ruan2016}; \citealt{MacLeold2016}; \citealt{Runnoe2016}; \citealt{McElroy2016}; \citealt{Gezari2017}). Possible explanations for this phenomenon include changes in accretion rate (e.g., \citealt{Marin2017}). On longer timescales, ${>}10^4\,{\rm yr}$, AGN variability is probed through extended AGN photoionised emission line regions (EELR) such as the {\it{Voorwerpjes}} (VP, e.g., \citealt{Lintott2009}; \citealt{Gagne2011}; \citealt{Keel2012}; \citealt{Keel2017}; \citealt{Sartori2016}; \citealt{Sartori2018a}) or ionisation and radio structures in our own Milky Way (e.g., \citealt{Su2010}; \citealt{Bland2013}). Such timescales may be linked to dramatic changes in the AGN accretion state, as suggested by the viscous radial inflow timescales in standard thin accretion discs (e.g., \citealt{Shakura1973}; \citealt{LaMassa2015}), which might happen in an analogous way to X-ray binaries (e.g., \citealt{Sobolewska2011}). The variability features observed at different timescales, corresponding to modulation of AGN luminosity during the active phase of supermassive black holes (SMBH) growth, may then be superimposed to longer AGN phases. \cite{Schawinski2015} suggested that SMBHs can switch on and off 100-1000 times with typical AGN phases lasting ${\sim}10^5\,{\rm yr}$, as supported also by some models and statistical arguments (e.g. \citealt{Martini2003}; \citealt{Novak2011}; \citealt{Gabor2013}; \citealt{King2015}).

The variability features described above span many orders of magnitude both in amplitude and timescales. In this letter, we concentrate on the optical-UV emission from AGN accretion discs, which also illuminates gas far from the nucleus. Variability may therefore be mostly attributed to non-uniform SMBH fuelling, which can arise from physical mechanisms in place at different spatial scales. First, major galaxy mergers can trigger the fuelling of a gas reservoir through the central region (e.g., \citealt{Barnes1991}). Internal galaxy dynamics and  gas temperature can then affect how and when this gas enters the accretion disc (e.g., \citealt{Hopkins2006}). Finally, the conversion of gravitational potential to luminosity depends on physical properties of the disc (e.g., structure and viscosity), as well as the system's reaction to perturbations (e.g., \citealt{Shakura1973}).

In this letter we propose a simple phenomenological model to link and describe the variability features of actively accreting SMBHs observed at different timescales. We assume that the observed variability is mostly driven by changes in SMBH fuelling, which near the nucleus can lead to changes in accretion rate and Eddington ratio ($L/L_{\rm Edd}$). The link is motivated by general galaxy population properties, such as the $L/L_{\rm Edd}$ distribution function (ERDF) and the power spectral density (PSD) of the $L/L_{\rm Edd}$ evolution with time. This framework allows us to test if and how variability in individual AGN can be linked to and explained by the distribution of $L/L_{\rm Edd}$ among the galaxy population.

\section{Framework and model}\label{sec:frame}

AGN variability is generally considered to be a stochastic process (e.g., \citealt{Kelly2011}). Therefore, every light curve can be interpreted as a realisation of an underlying set of statistical properties. In this context, \cite{Emmanoulopoulos2013} proposed an algorithm to generate light curves based on a probability density function (PDF) describing the distribution of fluxes, and a power spectral density (PSD) describing the distribution of time frequencies.

\begin{figure}
\includegraphics[width=0.47\textwidth]{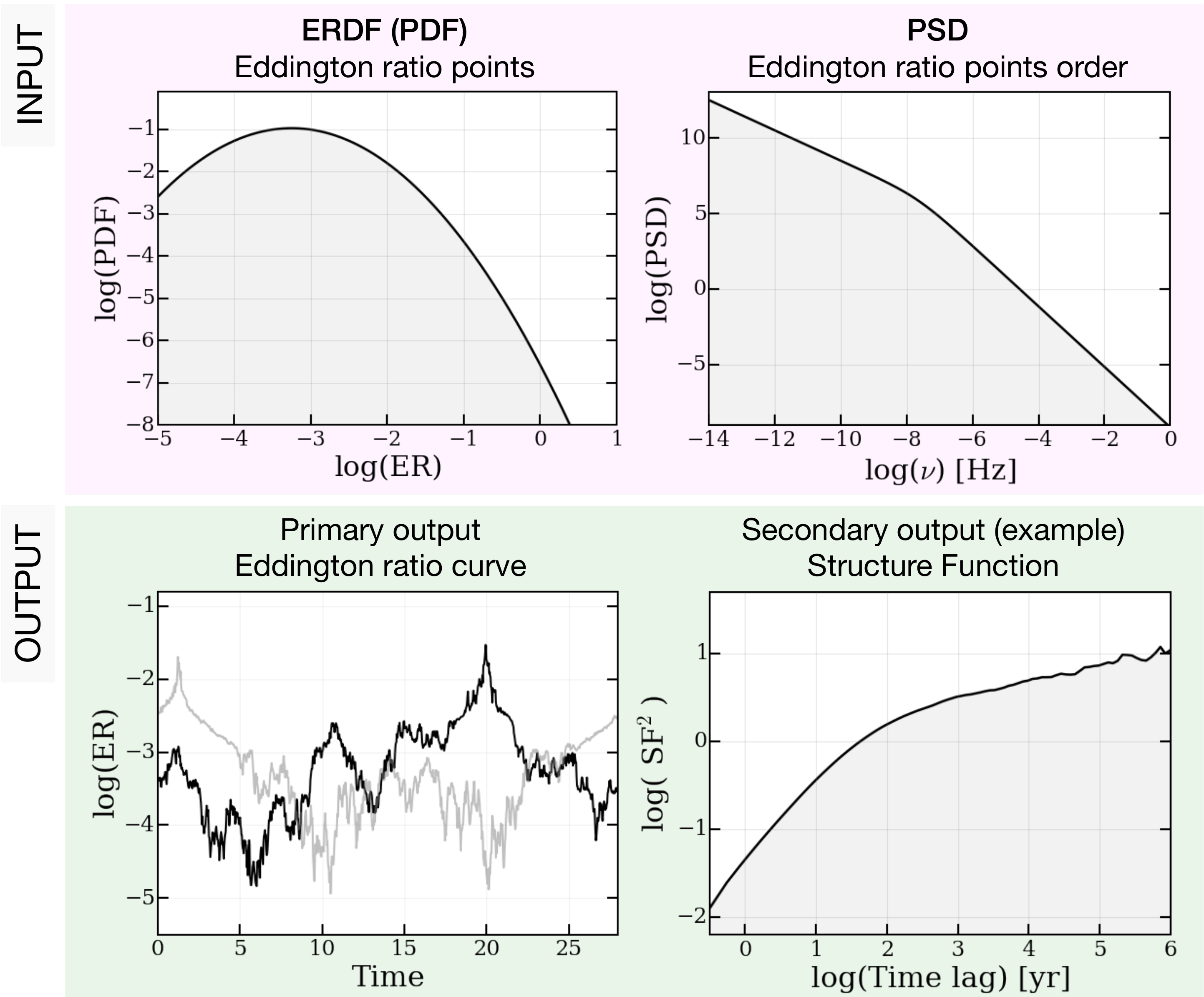}
\caption{Schematic summary of the $L/L_{\rm Edd}$ curve simulations. The input quantities, ERDF and PSD, provide information about the $L/L_{\rm Edd}$ values allowed in the $L/L_{\rm Edd}$ curve (ERDF) and how these points are ordered (PSD). The output $L/L_{\rm Edd}$ curve (primary output; grey and black lines correspond to two different realisations) is then converted into a light curve and used to produce observables to be compared to observations (secondary output, e.g., structure function). Plots are for illustrative purpose only.}
\label{fig:schem}
\end{figure}

Motivated by these ideas, we propose that AGN variability due to changes in SMBH fuelling can be modeled starting from the PDF and the PSD of the $L/L_{\rm Edd}$ evolution with time\footnote{In the following, we will refer to the time series representing the $L/L_{\rm Edd}$ evolution with time as ``$L/L_{\rm Edd}$ curve''.}. A summary of the proposed approach is illustrated in Fig. \ref{fig:schem}. By assuming that over long enough time periods every AGN should span the same $L/L_{\rm Edd}$ range as a static snapshot of the whole AGN population, the PDF shape may be inspired by the ERDF. On the other hand, the PSD of the $L/L_{\rm Edd}$ curve is likely to have a broken power-law shape, similar to what is observed for the light curves on timescales of hours to years (although we note that such light curves are expressed in magnitude). The bending may also be expected since variability power cannot increase indefinitely, as accretion is bounded by physical processes. Following the algorithm in \cite{Emmanoulopoulos2013}, the input PSD and PDF are used to generate $L/L_{\rm Edd}$ curves which, assuming a radiative efficiency and a description for the BH mass growth, are converted into light curves. Following a forward modelling approach, the light curves can then be used to compute observables (e.g., SF points or magnitude differences) to be compared to real observations. 

In this letter we provide a simplified proof of concept, motivated by and compared to some data available in the literature (Section\,\ref{sec:data}). We will present an elaborate investigation of this model, including numerical and other tests, as well as extensive comparison with data, in a forthcoming publication. Our hypothesis is that a single ERDF+PSD set, or a limited number of them, should be able to reproduce observed light curves (in a statistical sense), that are consistent with the variability features observed both at short ($\sim$yr) and long (${>}10^4\,{\rm yr}$) timescales. This simple model effectively links the variability of individual AGN to the underlying, and more accessible, properties of the entire population.


\section{Proof of concept}\label{sec:data} 

To show how our model can be used to link and describe AGN variability on different timescales, we compile a set of variability measurements in terms of magnitude differences ($\Delta m$) at a given time lag, motivated by the definition of the structure function (SF). In this work, the SF$^2$ of the light curve $m(t_i)$ at a time lag $\tau$ is defined as:

\begin{equation}\label{eq:SF_2}
{\rm SF}^2(\tau) = \frac{1}{P} \sum_{i,j> i} \left[ m(t_i) - m(t_j) \right] ^ 2 = \langle \left[ m(t) - m(t + \tau) \right] ^ 2\rangle
\end{equation}

\noindent
where $P$ is the number of magnitude pairs $\{m(t_i), m(t_j)\}$ with $t_j - t_i = \tau$, and the units are $[\rm mag ^2]$. The SF can be computed directly from a single AGN light curve, or using an ensemble approach (e.g., \citealt{MacLeod2010}). From Eq. \ref{eq:SF_2}, (${\Delta} m$)$^2$ measurements for single objects at given timescales can be directly compared to SF$^2$ values. This formalism therefore allows us to compare, on a single plot, the general AGN variability behaviour at different timescales, as well as with specific variability features (e.g., CLQSOs).

\begin{figure*}
\includegraphics[width=\textwidth]{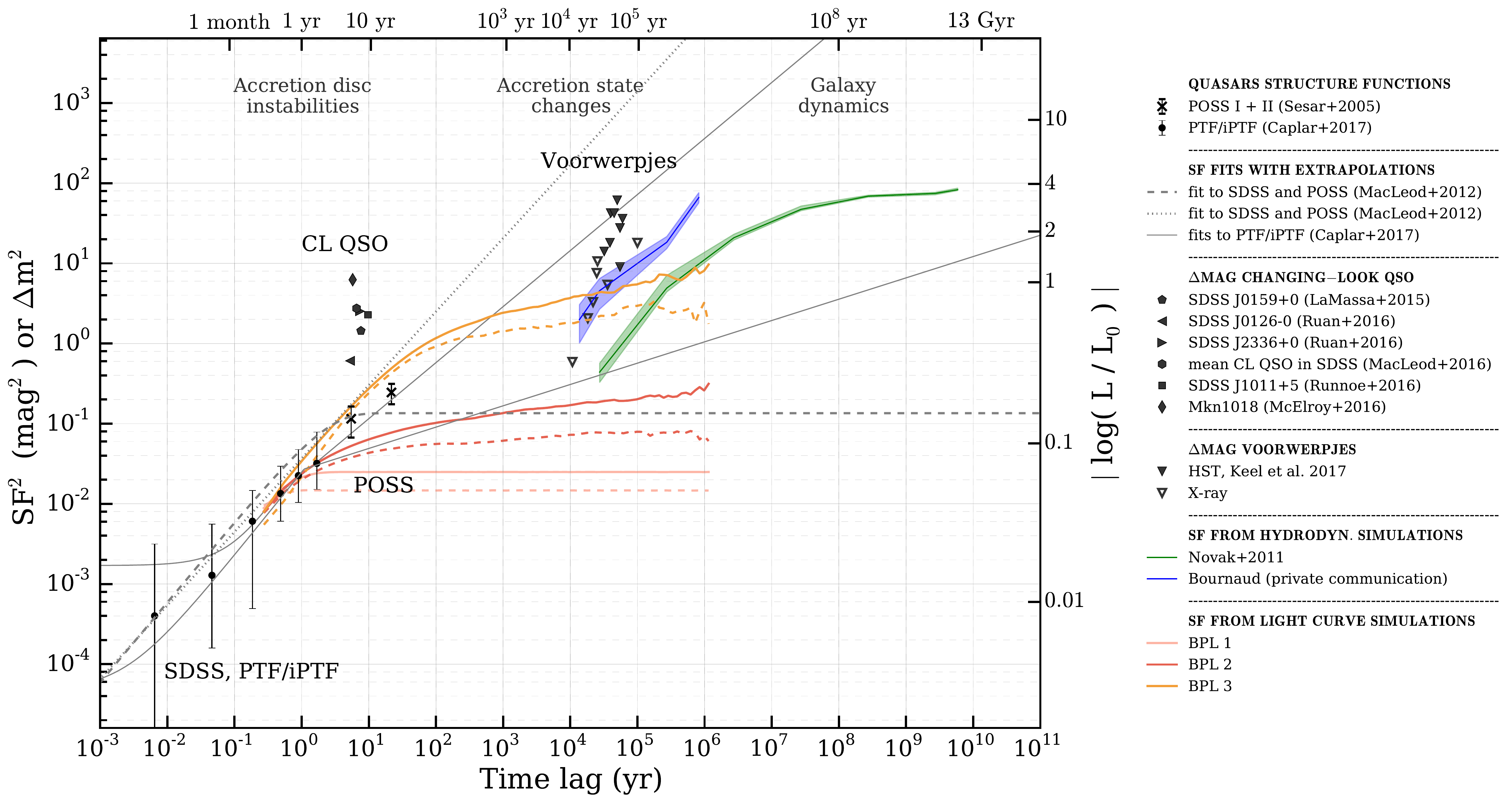}
\caption{Rest-frame SF and ${\Delta} m$ plot summarising variability data from the literature and from our own estimates (black points), fits to the SF points (grey lines) as well as simulations results (coloured lines). The SF from light curve simulations are computed both considering all the simulated points (solid line) or only the points with $L/L_{\rm Edd}$ $> 10^{-3}$ (dashed line) to mimic observational biases. See Section \ref{sec:data} for details and references.}
\label{fig:SF_plot}
\end{figure*}

In the following we describe how we compiled the rest-frame SF and ${\Delta} m$ measurements shown in Fig. \ref{fig:SF_plot}. We stress that all the measurements refer to galaxies currently hosting {\it{actively accreting}} SMBH (although EELR may in principle allow to find galaxies hosting currently inactive SMBHs), and that the reported time lags correspond to the time probed by the observations, as opposed to model predictions or timescales inferred from observed occurrence rates. Since sources generally exhibit chromatic variability properties (e.g., \citealt{VandenBerk2004}), it would be preferable to compare SF and ${\Delta} m$ in the same band. However, since the goal of the present work is to compile a wide range of observations covering different timescales we are forced to consider measurements obtained for various spectral regimes (although we mainly concentrate on optical and UV data, which are linked to accretion disc physics and SMBH fuelling). This will not affect our main findings, as the differences among optical and UV bands are generally lower than the level of precision required for our current analysis.

\subsection{Data compilation}

\subsubsection{Ensemble quasar variability}

Ensemble SFs for time lags between days and multiple decades  are often described as a damped random walk, an exponential or a broken power law (e.g. \citealt{MacLeod2012}; \citealt{Caplar2017}). In Fig. \ref{fig:SF_plot} we show the $r$-band SF points from the PTF/iPTF survey (\citealt{Caplar2017}) and the $g$-band SF points from POSS (\citealt{Sesar2006}). We also overplot example fits to the PTF/iPTF SF points (\citealt{Caplar2017}) and to SDSS and POSS (\citealt{MacLeod2012}) with extrapolation to longer timescales. When needed we converted SF measurements and fits to agree with the formulation in Eq. \ref{eq:SF_2}.

\subsubsection{Changing look QSOs}

We obtained ${\Delta} m$ values for single CLQSO from the literature, starting from the reported changes in magnitude or accretion rate (\citealt{LaMassa2015}; \citealt{Ruan2016}; \citealt{Runnoe2016}; \citealt{McElroy2016}). In addition we computed a mean (${\Delta} m$)$^2$ value for the SDSS CLQSOs in \citealt{MacLeold2016}. For the time lags we considered the shortest reported time difference between observations that bracket the AGN class change (as defined in the correspondent literature), converted into rest-frame. We note that some of these timescales may be upper limits.

\subsubsection{Extended emission line regions - Voorwerpjes}

We computed the ${\Delta} m$ values for the VP galaxies in two different ways. For 8 galaxies with available narrow-band {\it{HST}} imaging (\citealt{Keel2015}) we considered the difference between the observed nuclear emission and that required to account for the AGN-like emission in the most distant, AGN-photoionised region (assuming a constant spectral energy distribution shape; see  \citealt{Keel2017} for the ionisation history reconstruction). For 7 galaxies we estimated the present day AGN luminosities from the fit to the {\it{XMM-Newton}} and {\it{NuSTAR}} X-ray spectra and a conversion factor to bolometric luminosity (\citealt{Marconi2004}), and compared them to the past luminosities inferred from the ionising luminosity reported in \cite{Keel2012} (see also \citealt{Sartori2018a} for a similar analysis on IC 2497). The timescales therefore correspond to the difference between the times probed by the observations. We note that since the measured distances are projected distances, the reported time lags are lower limits. Because of the complexity of photoionisation physics and bolometric corrections, and the uncertainties in the travel time estimation, we caution that the obtained ${\Delta} m$ values and time lags have to be treated as order of magnitude estimates. We also stress that, because of the sample selection, the VP sample is biased towards AGN exhibiting large luminosity drops..



\subsubsection{Hydrodynamic simulations}

Most simulations that trace SMBH growth provide a history of BH accretion rates, which in turn can be converted into (bolometric) light curves. By assuming that the bolometric luminosity varies in the same way as the optical luminosity, the simulated light curves can be used to compute SFs to be compared with observations. In Fig. \ref{fig:SF_plot} we show the SFs obtained from the simulations in \cite{Novak2011} and from F. Bournaud (private communication). We note that simulations are currently the only possible way to investigate AGN variability on timescales longer than $10^5$ yr.

\subsection{Example model}\label{sec:model}

To illustrate a possible application of our model we simulate $L/L_{\rm Edd}$ curves with three different input PSD and PDF, and compute their SF in magnitude space. The obtained SF can then be compared to the values in Fig. \ref{fig:SF_plot} to test if the input PSD and PDF can explain the observed variability.

We chose a lognormal PDF with the same parameters as the ERDF proposed in \cite{Weigel2017}. For the PSD we considered three different broken power-laws with different slopes and break frequency (BPL 1: $\alpha_{\rm low} = 0$, $\alpha_{\rm high} = -2$, $\nu_{\rm break} = 10^{-8}$Hz $\sim$3yr; BPL 2: $\alpha_{\rm low} = -1$, $\alpha_{\rm high} = -2$, $\nu_{\rm break} = 10^{-8}$Hz $\sim$3yr; BPL 3: $\alpha_{\rm low} = -1$, $\alpha_{\rm high} = -2$, $\nu_{\rm break} = 2 \times 10^{-10}$Hz $\sim$160yr). We then assumed a linear conversion between $L/L_{\rm Edd}$ and luminosity (M$_{\rm BH}$ is not expected to increase significantly during the considered timescales) and computed the SF in magnitude space with Eq. \ref{eq:SF_2}. The resulting SF, arbitrarily renormalised to be consistent with the PTF/iPTF SF, are shown in Fig. \ref{fig:SF_plot}. The  origin of the normalisation mismatch may arise e.g. from numerical effects and/or fundamental physics (see below) and will be addressed in a future work.

We caution that because of computational limits and uncertainties due e.g. to the conversion from $L/L_{\rm Edd}$ to luminosity\footnote{The conversion from $L/L_{\rm Edd}$ to either bolometric or monochromatic luminosities is not fully understood yet, especially at low $L/L_{\rm Edd}$ regimes (ER $\ll 0.01$) where we expect a change from radiative efficient to radiative inefficient accretion (e.g., \citealt{Ho2009}).}, these example simulations have illustrative purposes only. The details of the simulation approach, as well as the comparison with the data, will be addressed in detail in a forthcoming publication. The aim of the current proof of concept is not to fit the data, but rather to qualitatively illustrate what different PSDs lead to.

\section{Discussion}\label{sec:conc} 

\subsection{Insights and results from the structure function plot}\label{sec:res}

The SF$^{2}$ and (${\Delta} m$)$^{2}$ plot in Fig. \ref{fig:SF_plot} summarises all the observed variability measurements described in Section \ref{sec:data}. The SF corresponding to the short timescales probed by ensemble analysis implies an increase in variability with increasing time lags, and hints at a flattening at $\tau{\sim}10^1 - 10^2$ yr. Extrapolating the fits to these SF measurements to longer timescales leads to disagreement. They are therefore not sufficient to predict variability at longer, ${>}10^4\,{\rm yr}$, timescales. 

The position of the VP galaxies on the plot provides further support for a flattening of the SF at time lags exceeding the regime probed by ensemble analysis. We stress that some of the VP galaxies were preselected to show large luminosity drops. Since VP are rare among general AGN population, the ensemble SF at these timescales may be much lower. The {\it{observed}} flattening of the SF is also a natural consequence of the fact that, at long time lags, a single power law SF would predict ${\Delta} m$ that are too large to be observed due to sensitivity limits and AGN selection criteria. Indeed, AGN with very low $L/L_{\rm Edd}$ 
($\sim$$10^{-4}$ or lower) are almost undetectable with current facilities, unless they are found in nearby galaxies with high quality data. Such selection biases can be included in our modeling approach. We will therefore be able to test if a single ERDF+PSD set, or a limited number of them, can reproduce the observed flattening and variability features.

The (${\Delta} m$)$^2$ values observed in some CLQSOs are at least one order of magnitude higher than the mean values given by the SF measured using the ensemble approach, in agreement with the results reported in many of the aforementioned CLQSO studies. Similar to what is observed for the VP, CLQSO are likely rare based on large searches (\citealt{Ruan2016}; \citealt{MacLeold2016}) and possibly related to other processes with higher variability. Modeling the AGN light curves and the (${\Delta} m$)$^2$ distributions may help understand if these objects are the extreme tail of the AGN phenomena, or if they represent a separate process (see \citealt{Rumbaugh2017}, \citealt{Graham2017} for a discussion about extreme variability in QSO).

The variability properties predicted by hydrodynamic simulations are strongly dependent on their spatial resolution and on the applied accretion and feedback recipes. This is consistent with the fact that the BH fueling, and thus variability in emission, is ultimately dominated by circumnuclear processes, on scales that are seldom addressed by simulations. Current hydrodynamic simulations are therefore not suited to investigate AGN variability. Fig. \ref{fig:SF_plot} illustrates the challenges that simulations are facing (e.g., \citealt{Negri2017}), and provides an additional way to test how closely future simulations will reflect observations.



\subsection{Insights from modeling}

In Section \ref{sec:model} we outlined an example application of our method. We considered three different ERDF+PSD sets. The first results suggest that a broken power-law shaped PSD may be able to explain the gross variability features observed. However, further investigation --- e.g., of the input definition and conversion from $L/L_{\rm Edd}$ curves to observables --- is needed to get more constraining results. This will be addressed in detail in a forthcoming paper.

We hypothesize that a single ERDF+PSD set, or a limited number of them, should be able to produce $L/L_{\rm Edd}$ curves, and therefore accretion rate and light curves, which are consistent with the observed variability features both at short ($\sim$yr) and long (${>}10^4\,{\rm yr}$) timescales. This assumption may be justified by the fact that the ERDF appears to be universal (\citealt{Weigel2017}), and by the BH accretion physics (e.g., \citealt{Shakura1973}). Since there is no obvious reason that accretion disc physics should evolve with time (the same physics is expected in local galaxies and in z\,$\sim$\,7 quasars), the short timescale variations connected to the accretion disc ought to be redshift independent. On the other hand, the long-timescale variability linked to cosmology may evolve with redshift (e.g., redshift-dependent gas accretion or merger rate, \citealt{Lotz2011}). In any case, since the observations in Fig. \ref{fig:SF_plot} are mostly for luminous quasars in low to intermediate redshift galaxies, we cannot probe the evolution with redshift with the current data. We also note that, as suggested by some ensemble studies (e.g. \citealt{Caplar2017}; \citealt{Rumbaugh2017}), short timescale variability could depend on $L/L_{\rm Edd}$, luminosity and black hole mass M$_{\rm BH}$. However, the data analysed here do not seem to require a model dependent on these quantities. We will investigate this aspect further in future studies.

If our hypothesis is correct, the underlying PSD would be the temporal analogue in AGN physics to the matter power spectrum in cosmology: similarly to how a single matter power spectrum is responsible for multiple matter assemblages (e.g. \citealt{Frenk2012}), from large cosmic structures down to dwarf galaxies, the AGN power spectrum may be responsible for AGN variability over many orders of magnitude in time. If this is true, our model will allow us to link and describe AGN variability at different timescales. Our framework will allow us to test this hypothesis and to constrain the shape of the underlying PSD, providing novel insights into the AGN variability phenomenon.

\section*{Acknowledgements}

We acknowledge support from SNSF Grants PP00P2\_138979 and PP00P2\_166159 (LS, KS); NASA through ADAP award NNH16CT03C (MJK); CONICYT grant Basal-CATA PFB-06/2007 and FONDECYT Regular 1160999 (ET).




\begin{thebibliography}{}
\makeatletter
\relax
\def\mn@urlcharsother{\let\do\@makeother \do\$\do\&\do\#\do\^\do\_\do\%\do\~}
\def\mn@doi{\begingroup\mn@urlcharsother \@ifnextchar [ {\mn@doi@}
  {\mn@doi@[]}}
\def\mn@doi@[#1]#2{\def\@tempa{#1}\ifx\@tempa\@empty \href
  {http://dx.doi.org/#2} {doi:#2}\else \href {http://dx.doi.org/#2} {#1}\fi
  \endgroup}
\def\mn@eprint#1#2{\mn@eprint@#1:#2::\@nil}
\def\mn@eprint@arXiv#1{\href {http://arxiv.org/abs/#1} {{\tt arXiv:#1}}}
\def\mn@eprint@dblp#1{\href {http://dblp.uni-trier.de/rec/bibtex/#1.xml}
  {dblp:#1}}
\def\mn@eprint@#1:#2:#3:#4\@nil{\def\@tempa {#1}\def\@tempb {#2}\def\@tempc
  {#3}\ifx \@tempc \@empty \let \@tempc \@tempb \let \@tempb \@tempa \fi \ifx
  \@tempb \@empty \def\@tempb {arXiv}\fi \@ifundefined
  {mn@eprint@\@tempb}{\@tempb:\@tempc}{\expandafter \expandafter \csname
  mn@eprint@\@tempb\endcsname \expandafter{\@tempc}}}

\bibitem[\protect\citeauthoryear{{Barnes} \& {Hernquist}}{{Barnes} \&
  {Hernquist}}{1991}]{Barnes1991}
{Barnes} J.~E.,  {Hernquist} L.~E.,  1991, \mn@doi [\apjl] {10.1086/185978},
  \href {http://adsabs.harvard.edu/abs/1991ApJ...370L..65B} {370, L65}

\bibitem[\protect\citeauthoryear{{Bland-Hawthorn}, {Maloney}, {Sutherland}  \&
  {Madsen}}{{Bland-Hawthorn} et~al.}{2013}]{Bland2013}
{Bland-Hawthorn} J.,  {Maloney} P.~R.,  {Sutherland} R.~S.,   {Madsen} G.~J.,
  2013, \mn@doi [\apj] {10.1088/0004-637X/778/1/58}, \href
  {http://adsabs.harvard.edu/abs/2013ApJ...778...58B} {778, 58}

\bibitem[\protect\citeauthoryear{{Caplar}, {Lilly}  \& {Trakhtenbrot}}{{Caplar}
  et~al.}{2017}]{Caplar2017}
{Caplar} N.,  {Lilly} S.~J.,   {Trakhtenbrot} B.,  2017, \mn@doi [\apj]
  {10.3847/1538-4357/834/2/111}, \href
  {http://adsabs.harvard.edu/abs/2017ApJ...834..111C} {834, 111}

\bibitem[\protect\citeauthoryear{{Emmanoulopoulos}, {McHardy}  \&
  {Papadakis}}{{Emmanoulopoulos} et~al.}{2013}]{Emmanoulopoulos2013}
{Emmanoulopoulos} D.,  {McHardy} I.~M.,   {Papadakis} I.~E.,  2013, \mn@doi
  [\mnras] {10.1093/mnras/stt764}, \href
  {http://adsabs.harvard.edu/abs/2013MNRAS.433..907E} {433, 907}

\bibitem[\protect\citeauthoryear{{Frenk} \& {White}}{{Frenk} \&
  {White}}{2012}]{Frenk2012}
{Frenk} C.~S.,  {White} S.~D.~M.,  2012, \mn@doi [Annalen der Physik]
  {10.1002/andp.201200212}, \href
  {http://adsabs.harvard.edu/abs/2012AnP...524..507F} {524, 507}

\bibitem[\protect\citeauthoryear{{Gabor} \& {Bournaud}}{{Gabor} \&
  {Bournaud}}{2013}]{Gabor2013}
{Gabor} J.~M.,  {Bournaud} F.,  2013, \mn@doi [\mnras] {10.1093/mnras/stt1046},
  \href {http://adsabs.harvard.edu/abs/2013MNRAS.434..606G} {434, 606}

\bibitem[\protect\citeauthoryear{{Gagne}, {Crenshaw}, {Keel}  \&
  {Fischer}}{{Gagne} et~al.}{2011}]{Gagne2011}
{Gagne} J.,  {Crenshaw} D.~M.,  {Keel} W.~C.,   {Fischer} T.~C.,  2011, in
  American Astronomical Society Meeting Abstracts \#217. p. 142.12

\bibitem[\protect\citeauthoryear{{Gezari} et~al.,}{{Gezari}
  et~al.}{2017}]{Gezari2017}
{Gezari} S.,  et~al., 2017, \mn@doi [\apj] {10.3847/1538-4357/835/2/144}, \href
  {http://adsabs.harvard.edu/abs/2017ApJ...835..144G} {835, 144}

\bibitem[\protect\citeauthoryear{{Graham}, {Djorgovski}, {Drake}, {Stern},
  {Mahabal}, {Glikman}, {Larson}  \& {Christensen}}{{Graham}
  et~al.}{2017}]{Graham2017}
{Graham} M.~J.,  {Djorgovski} S.~G.,  {Drake} A.~J.,  {Stern} D.,  {Mahabal}
  A.~A.,  {Glikman} E.,  {Larson} S.,   {Christensen} E.,  2017, \mn@doi
  [\mnras] {10.1093/mnras/stx1456}, \href
  {http://adsabs.harvard.edu/abs/2017MNRAS.470.4112G} {470, 4112}

\bibitem[\protect\citeauthoryear{{Ho}}{{Ho}}{2009}]{Ho2009}
{Ho} L.~C.,  2009, \mn@doi [\apj] {10.1088/0004-637X/699/1/626}, \href
  {http://adsabs.harvard.edu/abs/2009ApJ...699..626H} {699, 626}

\bibitem[\protect\citeauthoryear{{Hopkins} \& {Hernquist}}{{Hopkins} \&
  {Hernquist}}{2006}]{Hopkins2006}
{Hopkins} P.~F.,  {Hernquist} L.,  2006, \mn@doi [\apjs] {10.1086/505753},
  \href {http://adsabs.harvard.edu/abs/2006ApJS..166....1H} {166, 1}

\bibitem[\protect\citeauthoryear{{Keel} et~al.,}{{Keel}
  et~al.}{2012}]{Keel2012}
{Keel} W.~C.,  et~al., 2012, \mn@doi [mnras]
  {10.1111/j.1365-2966.2011.20101.x}, \href
  {http://adsabs.harvard.edu/abs/2012MNRAS.420..878K} {420, 878}

\bibitem[\protect\citeauthoryear{{Keel} et~al.,}{{Keel}
  et~al.}{2015}]{Keel2015}
{Keel} W.~C.,  et~al., 2015, \mn@doi [\aj] {10.1088/0004-6256/149/5/155}, \href
  {http://adsabs.harvard.edu/abs/2015AJ....149..155K} {149, 155}

\bibitem[\protect\citeauthoryear{{Keel} et~al.,}{{Keel}
  et~al.}{2017}]{Keel2017}
{Keel} W.~C.,  et~al., 2017, \mn@doi [ApJ] {10.3847/1538-4357/835/2/256}, \href
  {http://adsabs.harvard.edu/abs/2017ApJ...835..256K} {835, 256}

\bibitem[\protect\citeauthoryear{{Kelly}, {Sobolewska}  \&
  {Siemiginowska}}{{Kelly} et~al.}{2011}]{Kelly2011}
{Kelly} B.~C.,  {Sobolewska} M.,   {Siemiginowska} A.,  2011, \mn@doi [\apj]
  {10.1088/0004-637X/730/1/52}, \href
  {http://adsabs.harvard.edu/abs/2011ApJ...730...52K} {730, 52}

\bibitem[\protect\citeauthoryear{{King} \& {Nixon}}{{King} \&
  {Nixon}}{2015}]{King2015}
{King} A.,  {Nixon} C.,  2015, \mn@doi [\mnras] {10.1093/mnrasl/slv098}, \href
  {http://adsabs.harvard.edu/abs/2015MNRAS.453L..46K} {453, L46}

\bibitem[\protect\citeauthoryear{{LaMassa} et~al.,}{{LaMassa}
  et~al.}{2015}]{LaMassa2015}
{LaMassa} S.~M.,  et~al., 2015, \mn@doi [\apj] {10.1088/0004-637X/800/2/144},
  \href {http://adsabs.harvard.edu/abs/2015ApJ...800..144L} {800, 144}

\bibitem[\protect\citeauthoryear{{Lintott} et~al.,}{{Lintott}
  et~al.}{2009}]{Lintott2009}
{Lintott} C.~J.,  et~al., 2009, \mn@doi [\mnras]
  {10.1111/j.1365-2966.2009.15299.x}, \href
  {http://adsabs.harvard.edu/abs/2009MNRAS.399..129L} {399, 129}

\bibitem[\protect\citeauthoryear{{Lotz}, {Jonsson}, {Cox}, {Croton}, {Primack},
  {Somerville}  \& {Stewart}}{{Lotz} et~al.}{2011}]{Lotz2011}
{Lotz} J.~M.,  {Jonsson} P.,  {Cox} T.~J.,  {Croton} D.,  {Primack} J.~R.,
  {Somerville} R.~S.,   {Stewart} K.,  2011, \mn@doi [\apj]
  {10.1088/0004-637X/742/2/103}, \href
  {http://adsabs.harvard.edu/abs/2011ApJ...742..103L} {742, 103}

\bibitem[\protect\citeauthoryear{{MacLeod} et~al.,}{{MacLeod}
  et~al.}{2010}]{MacLeod2010}
{MacLeod} C.~L.,  et~al., 2010, \mn@doi [\apj] {10.1088/0004-637X/721/2/1014},
  \href {http://adsabs.harvard.edu/abs/2010ApJ...721.1014M} {721, 1014}

\bibitem[\protect\citeauthoryear{{MacLeod} et~al.,}{{MacLeod}
  et~al.}{2012}]{MacLeod2012}
{MacLeod} C.~L.,  et~al., 2012, \mn@doi [\apj] {10.1088/0004-637X/753/2/106},
  \href {http://adsabs.harvard.edu/abs/2012ApJ...753..106M} {753, 106}

\bibitem[\protect\citeauthoryear{{MacLeod} et~al.,}{{MacLeod}
  et~al.}{2016}]{MacLeold2016}
{MacLeod} C.~L.,  et~al., 2016, \mn@doi [\mnras] {10.1093/mnras/stv2997}, \href
  {http://adsabs.harvard.edu/abs/2016MNRAS.457..389M} {457, 389}

\bibitem[\protect\citeauthoryear{{Marconi}, {Risaliti}, {Gilli}, {Hunt},
  {Maiolino}  \& {Salvati}}{{Marconi} et~al.}{2004}]{Marconi2004}
{Marconi} A.,  {Risaliti} G.,  {Gilli} R.,  {Hunt} L.~K.,  {Maiolino} R.,
  {Salvati} M.,  2004, \mn@doi [\mnras] {10.1111/j.1365-2966.2004.07765.x},
  \href {http://adsabs.harvard.edu/abs/2004MNRAS.351..169M} {351, 169}

\bibitem[\protect\citeauthoryear{{Marin}}{{Marin}}{2017}]{Marin2017}
{Marin} F.,  2017, \mn@doi [\aap] {10.1051/0004-6361/201731726}, \href
  {http://adsabs.harvard.edu/abs/2017A%26A...607A..40M} {607, A40}

\bibitem[\protect\citeauthoryear{{Martini} \& {Schneider}}{{Martini} \&
  {Schneider}}{2003}]{Martini2003}
{Martini} P.,  {Schneider} D.~P.,  2003, \mn@doi [\apjl] {10.1086/379888},
  \href {http://adsabs.harvard.edu/abs/2003ApJ...597L.109M} {597, L109}

\bibitem[\protect\citeauthoryear{{McElroy} et~al.,}{{McElroy}
  et~al.}{2016}]{McElroy2016}
{McElroy} R.~E.,  et~al., 2016, \mn@doi [\aap] {10.1051/0004-6361/201629102},
  \href {http://adsabs.harvard.edu/abs/2016A%26A...593L...8M} {593, L8}

\bibitem[\protect\citeauthoryear{{Negri} \& {Volonteri}}{{Negri} \&
  {Volonteri}}{2017}]{Negri2017}
{Negri} A.,  {Volonteri} M.,  2017, \mn@doi [\mnras] {10.1093/mnras/stx362},
  \href {http://adsabs.harvard.edu/abs/2017MNRAS.467.3475N} {467, 3475}

\bibitem[\protect\citeauthoryear{{Novak}, {Ostriker}  \& {Ciotti}}{{Novak}
  et~al.}{2011}]{Novak2011}
{Novak} G.~S.,  {Ostriker} J.~P.,   {Ciotti} L.,  2011, \mn@doi [\apj]
  {10.1088/0004-637X/737/1/26}, \href
  {http://adsabs.harvard.edu/abs/2011ApJ...737...26N} {737, 26}

\bibitem[\protect\citeauthoryear{{Ruan} et~al.,}{{Ruan}
  et~al.}{2016}]{Ruan2016}
{Ruan} J.~J.,  et~al., 2016, \mn@doi [\apj] {10.3847/0004-637X/826/2/188},
  \href {http://adsabs.harvard.edu/abs/2016ApJ...826..188R} {826, 188}

\bibitem[\protect\citeauthoryear{{Rumbaugh} et~al.,}{{Rumbaugh}
  et~al.}{2017}]{Rumbaugh2017}
{Rumbaugh} N.,  et~al., 2017, preprint, \href
  {http://adsabs.harvard.edu/abs/2017arXiv170607875R} {} (\mn@eprint {arXiv}
  {1706.07875})

\bibitem[\protect\citeauthoryear{{Runnoe} et~al.,}{{Runnoe}
  et~al.}{2016}]{Runnoe2016}
{Runnoe} J.~C.,  et~al., 2016, \mn@doi [\mnras] {10.1093/mnras/stv2385}, \href
  {http://adsabs.harvard.edu/abs/2016MNRAS.455.1691R} {455, 1691}

\bibitem[\protect\citeauthoryear{{Sartori} et~al.,}{{Sartori}
  et~al.}{2016}]{Sartori2016}
{Sartori} L.~F.,  et~al., 2016, \mn@doi [\mnras] {10.1093/mnras/stw230}, \href
  {http://adsabs.harvard.edu/abs/2016MNRAS.457.3629S} {457, 3629}

\bibitem[\protect\citeauthoryear{{Sartori} et~al.,}{{Sartori}
  et~al.}{2018}]{Sartori2018a}
{Sartori} L.~F.,  et~al., 2018, \mn@doi [\mnras] {10.1093/mnras/stx2952}, \href
  {http://adsabs.harvard.edu/abs/2018MNRAS.474.2444S} {474, 2444}

\bibitem[\protect\citeauthoryear{{Schawinski}, {Koss}, {Berney}  \&
  {Sartori}}{{Schawinski} et~al.}{2015}]{Schawinski2015}
{Schawinski} K.,  {Koss} M.,  {Berney} S.,   {Sartori} L.~F.,  2015, \mn@doi
  [\mnras] {10.1093/mnras/stv1136}, \href
  {http://adsabs.harvard.edu/abs/2015MNRAS.451.2517S} {451, 2517}

\bibitem[\protect\citeauthoryear{{Sesar} et~al.,}{{Sesar}
  et~al.}{2006}]{Sesar2006}
{Sesar} B.,  et~al., 2006, \mn@doi [\aj] {10.1086/503672}, \href
  {http://adsabs.harvard.edu/abs/2006AJ....131.2801S} {131, 2801}

\bibitem[\protect\citeauthoryear{{Shakura} \& {Sunyaev}}{{Shakura} \&
  {Sunyaev}}{1973}]{Shakura1973}
{Shakura} N.~I.,  {Sunyaev} R.~A.,  1973, \aap, \href
  {http://adsabs.harvard.edu/abs/1973A%26A....24..337S} {24, 337}

\bibitem[\protect\citeauthoryear{{Sobolewska}, {Siemiginowska}  \&
  {Gierli{\'n}ski}}{{Sobolewska} et~al.}{2011}]{Sobolewska2011}
{Sobolewska} M.~A.,  {Siemiginowska} A.,   {Gierli{\'n}ski} M.,  2011, \mn@doi
  [\mnras] {10.1111/j.1365-2966.2011.18302.x}, \href
  {http://adsabs.harvard.edu/abs/2011MNRAS.413.2259S} {413, 2259}

\bibitem[\protect\citeauthoryear{{Su}, {Slatyer}  \& {Finkbeiner}}{{Su}
  et~al.}{2010}]{Su2010}
{Su} M.,  {Slatyer} T.~R.,   {Finkbeiner} D.~P.,  2010, \mn@doi [\apj]
  {10.1088/0004-637X/724/2/1044}, \href
  {http://adsabs.harvard.edu/abs/2010ApJ...724.1044S} {724, 1044}

\bibitem[\protect\citeauthoryear{{Vanden Berk} et~al.,}{{Vanden Berk}
  et~al.}{2004}]{VandenBerk2004}
{Vanden Berk} D.~E.,  et~al., 2004, \mn@doi [\apj] {10.1086/380563}, \href
  {http://adsabs.harvard.edu/abs/2004ApJ...601..692V} {601, 692}

\bibitem[\protect\citeauthoryear{{Weigel}, {Schawinski}, {Caplar}, {Wong},
  {Treister}  \& {Trakhtenbrot}}{{Weigel} et~al.}{2017}]{Weigel2017}
{Weigel} A.~K.,  {Schawinski} K.,  {Caplar} N.,  {Wong} O.~I.,  {Treister} E.,
   {Trakhtenbrot} B.,  2017, \mn@doi [\apj] {10.3847/1538-4357/aa803b}, \href
  {http://adsabs.harvard.edu/abs/2017ApJ...845..134W} {845, 134}

\makeatother
\end{thebibliography}






\bsp	
\label{lastpage}
\end{document}